\documentclass[runningheads]{svmult}

\usepackage{makeidx}   % allows index generation
\usepackage{graphicx}  % standard LaTeX graphics tool
\usepackage{subeqnar}  % subnumbers individual equations
\usepackage{multicol}  % used for the two-column index
\usepackage{cropmark}  % cropmarks for pages without
\usepackage{physprbb}  % modified textarea for proceedings,
\newcommand{\greeksym}[1]{{\usefont{U}{psy}{m}{n}#1}}

\newcommand{\uDelta}{\mbox{\greeksym{D}}}

\newcommand{\uGamma}{\mbox{\greeksym{G}}}
\usepackage{lscape}

\begin{document}
\title*{Imaging Across the Spectrum: Synergies Between SKA and Other Future
  Telescopes}
\toctitle{Synergies Between SKA and Other Future Telescopes}
% allows explicit linebreak for the table of content
%
%
\titlerunning{Imaging across the spectrum}
% allows abbreviation of title, if the full title is too long
% to fit in the running head
%
\author{Andrei Lobanov}
\authorrunning{Lobanov}
% if there are more than two authors,
% please abbreviate author list for running head
%
%
\institute{Max-Planck-Institut f\"ur Radioastronomie, 
Auf dem H\"ugel 69, 53121 Bonn, Germany}

\maketitle              % typesets the title of the contribution

\begin{abstract}
  SKA\footnote{The Square Kilometer Array, a next generation interferometric
    instrument for centimeter-wave radio astronomy. See
    http://www.skatelescope.org for a detailed description.}  will be
  operating at the same time with several new large optical, X--ray and
  $\uGamma$--ray facilities currently under construction or planned. Fostering
  synergies in astrophysical research made across different spectral bands
  presents a compelling argument for designing the SKA such that it would
  offer imaging capabilities similar to those of other future telescopes.
  Imaging capabilities of the SKA are compared here with those of the major
  future astrophysical facilities.

\end{abstract}

\section{Imaging performance of SKA and other future telescopes}

A first order comparison of imaging performance of different instruments can
be made by comparing their respective spatial dynamic ranges (SDR, the ratio
between the maximum and the minimum detectable angular scales) and
resolutions. 

\begin{figure}[t]
\centerline{
\includegraphics[width=.42\textwidth,angle=-90,bb=0 0 595 700,clip=true]{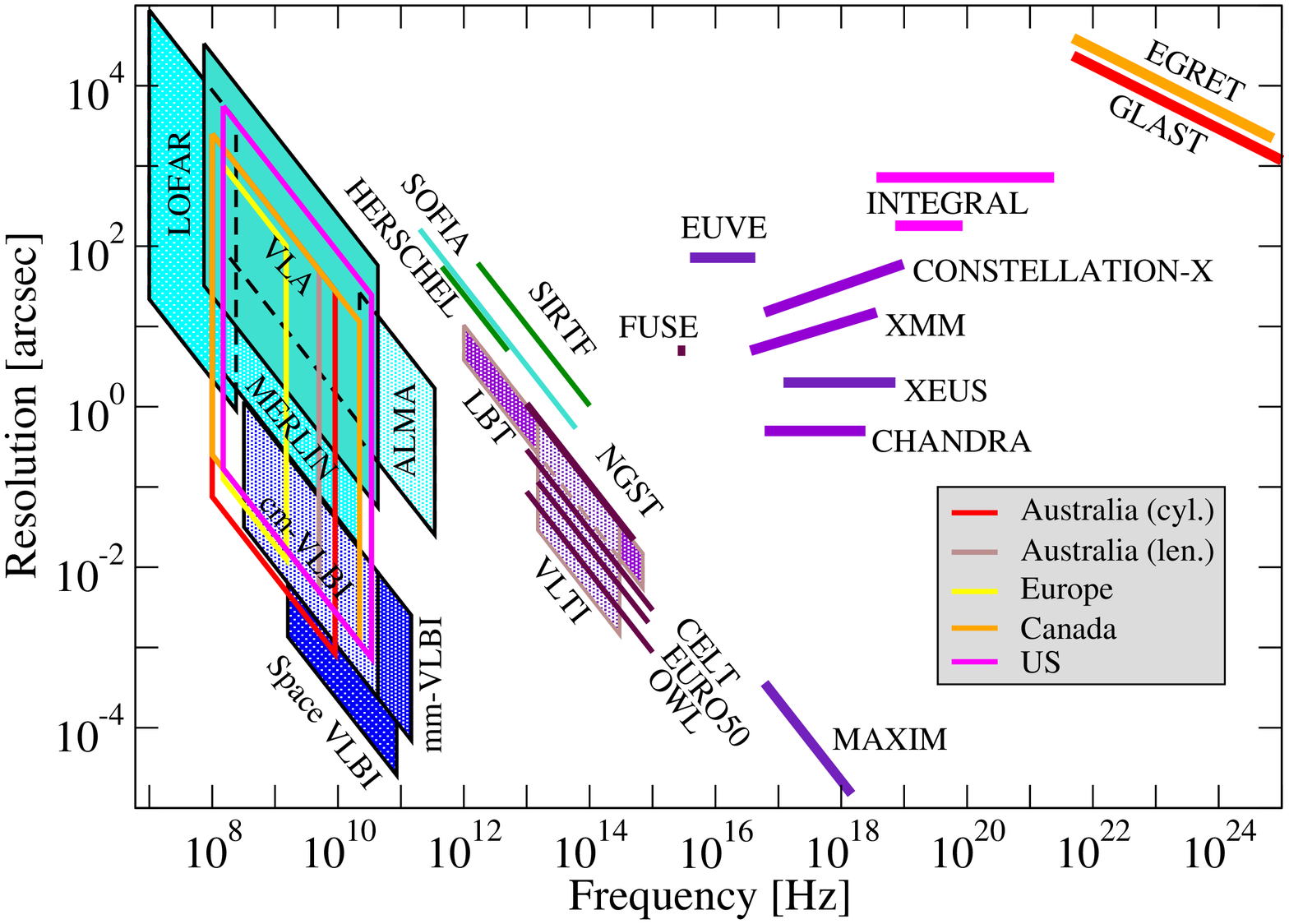}
\includegraphics[width=.42\textwidth,angle=-90,bb=0 0 595 700,clip=true]{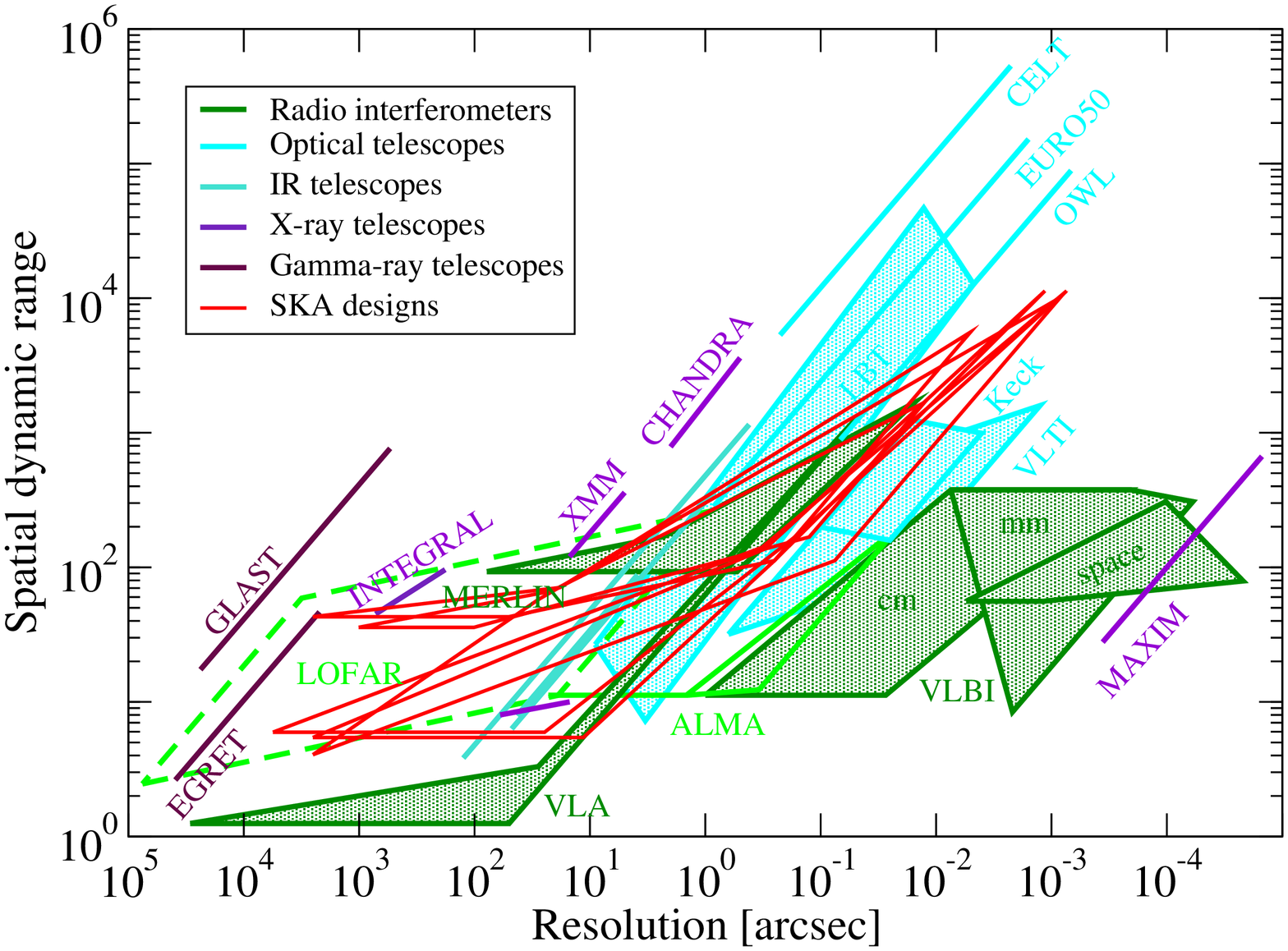}
}
\caption[]{{\bf Left:} Resolution of the SKA compared with the resolution of
  other main existing and future astronomical instruments. {\bf Right:}
Spatial dynamic range of the SKA designs for observations with $\uDelta
\nu$=1\,MHz and $\tau$=1\,s) compared to other major instruments~\cite{lobanov:ska}.}
\label{fg:lobanovF1}
\end{figure}

The resolution of the SKA is compared in Figure~\ref{fg:lobanovF1} with the
resolutions of various existing and future telescopes.  The resolution of SKA
is close to the resolution of the largest projected optical telescopes, but it
may be inferior to the resolution of the proposed X--ray interferometer
mission MAXIM. For the optical telescopes, the SKA would be able to present a
reasonable match provided that the SKA reaches a $\le 1$\,mas resolution at
its highest observing frequency (see~\cite{lobanov:ska} for more details).

The SDR of the different SKA designs is compared in Figure~\ref{fg:lobanovF1}
with the SDR of other instruments.  For a radio interferometer, the SDR is
affected by the observing bandwidth, $\uDelta\nu$, averaging time, $\tau$, and
filling factor of the Fourier domain, $\uDelta u/u$~\cite{lobanov:ska} (analogies of these
quantities can be found for all instruments working in other spectral bands).

For most of the high--dynamic range observations with the SKA the
bandwidth and integration time may have to be significantly reduced,
if one would require to reach SDR similar to that of the largest
optical instruments. These two corrections can be introduced at the
stage of observation preparation, and their worst effect is the
increased observing time needed to reach the required sensitivity. The
Fourier space sampling, described by the $\uDelta u/u$ ratio, is
however ``hard--wired'' into the array design, and can only be
improved by adding new stations. 
For an inhomogeneous array in which $\uDelta
u/u$ varies depending on the baseline length, the reduction of spatial
and even conventional dynamic range may be substantial.
This dictates the need to
optimize this parameter at the earliest possibles stages of the array
design.  In addition to that, optimization of $\uDelta u/u$ is also
required by high--fidelity imaging at low SNR levels. The lowest SNR
of ``trustable'' pixel in an interferometric image is given by
\begin{equation}
\ln(\rm SNR_{\rm low}) = \left[\frac{\pi}{4}\left(\frac{\uDelta u }{u} +1\right)\right]^2
\frac{1}{\ln 2}\,.
\end{equation}
The SDR reduction due to poor Fourier space sampling becomes significant at
$\uDelta u/u \ge 0.4$, and it is negligible at $\uDelta u/u \le 0.2$. It should
therefore be possible to reach the maximum SDR levels in an array
configuration that provides $\uDelta u /u \le 0.2$ at all baselines.  If
multifrequency synthesis is used for imaging, this condition becomes $\uDelta
u/u \le 0.2 + \uDelta\nu_\mathrm{mfs}$ ($\uDelta\nu_\mathrm{mfs}$ is the
fractional bandwidth over which the synthesis is performed).  Therefore, this
requirement must be considered as one of the basic requirements for the design
of the SKA.

To make the SKA a competitive imaging instrument that would match the
capabilities of future optical and X-ray telescopes, two
basic conditions must be fulfilled:
\begin{itemize}
\item[1.] Resolution of $\le 1$\,mas at the highest observing frequency.
\item[2.] Fourier plane filling factor $\uDelta u/u\le 0.2$ over the entire range of {\it uv}--coverage.
\end{itemize}

%INDEX%%%%%%%%%%%%%%%%%%%%%%%%%%%%%%%%%%%%%%%%%%%%%%%%%%%%%%%%%%%%%%%
% Please check with the editor of your book whether he plans to
% include a "mutual" subject index - if so, please code your entries
% in the standard syntax. For your own purposes you may print your
% "personal" index by using the following commands:
%
%\clearpage
%\addcontentsline{toc}{section}{Index}
%\flushbottom
%\printindex
%%%%%%%%%%%%%%%%%%%%%%%%%%%%%%%%%%%%%%%%%%%%%%%%%%%%%%%%%%%%%%%%%%%%%

\end{document}